\documentclass[11pt,twocolumn]{article}

\usepackage[T1]{fontenc}
\usepackage[utf8]{inputenc}
\usepackage{microtype}
\usepackage{booktabs}
\usepackage{tabularx}
\usepackage{multirow}
\usepackage{graphicx}
\usepackage{xcolor}
\usepackage{xurl}
\usepackage{hyperref}
\hypersetup{
  colorlinks = true,
  citecolor  = {blue!55!black},
  linkcolor  = {blue!55!black},
  urlcolor   = {blue!55!black},
}
\usepackage{natbib}
\usepackage{amsmath}
\usepackage{listings}
\usepackage{tikz}
\usetikzlibrary{positioning, fit, backgrounds, arrows.meta,
                decorations.pathreplacing}
\usepackage{geometry}
\lstdefinestyle{x402}{
  basicstyle=\ttfamily\footnotesize,
  breaklines=true,
  breakatwhitespace=false,
  frame=single,
  framerule=0.4pt,
  rulecolor=\color{gray!50},
  backgroundcolor=\color{gray!6},
  xleftmargin=4pt,
  xrightmargin=4pt,
  aboveskip=4pt,
  belowskip=4pt,
}

\newcommand{\sys}{\textsc{HardenedX402Client}}
\newcommand{\pkg}{\texttt{presidio-hardened-x402}}
\newcommand{\F}[1]{\texttt{#1}}

\newcommand{\circled}[1]{\tikz[baseline=(c.base)]{%
  \node[circle, inner sep=1.2pt, draw=gray!70, fill=gray!10,
        font=\tiny\sffamily] (c) {#1};}}

\title{Hardening x402: PII-Safe Agentic Payments\\via Pre-Execution Metadata Filtering}

\author{Vladimir Stantchev\\
  \small SRH University Heidelberg, Germany\\
  \small PRESIDIO Group, Sofia, Bulgaria\\
  \small \texttt{vladimir.stantchev@srh.de} \quad \texttt{vladimir@presidio-group.eu}
}

\date{April 2026}

\begin{document}
\maketitle

%% ============================================================
\begin{abstract}
  AI agents that pay for resources via the x402 protocol embed payment
  metadata --- resource URLs, descriptions, and reason strings --- in
  every HTTP payment request. This metadata is transmitted to the payment
  server and to the centralised facilitator API before any on-chain
  settlement occurs; neither party is typically bound by a data processing
  agreement. We present \pkg{},
  the first open-source middleware that intercepts x402 payment requests
  \emph{before} transmission to detect and redact personally identifiable
  information (PII), enforce declarative spending policies, and block
  duplicate replay attempts. To evaluate the PII filter, we construct a
  labeled synthetic corpus of 2,000 x402 metadata triples spanning seven
  use-case categories, and run a 42-configuration precision/recall sweep
  across two detection modes (regex, NLP) and five confidence thresholds.
  The recommended configuration (\texttt{mode=nlp}, \texttt{min\_score=0.5},
  all entity types) achieves micro-F1 = 0.898 with precision 0.972, at a
  p99 latency of 5.73\,ms --- well within the 50\,ms overhead budget.
  The middleware, corpus, and all experiment code are publicly available at
  \url{https://github.com/presidio-v/presidio-hardened-x402}.
\end{abstract}

%% ============================================================
\section{Introduction}
\label{sec:intro}

A broker who handles your money also handles your metadata. In x402,
every payment request carries three strings --- a resource URL, a
description, a reason --- transmitted first to the server that charges
the agent, then to the centralised facilitator API that settles the
payment. Neither party advertises a data retention policy. Neither party
is typically bound by a data processing agreement with the agent's
operator. The strings may contain a name, an email address, a social
security number. The infrastructure was designed for financial settlement,
not for privacy.

The x402 protocol~\cite{coinbase2024x402} is an HTTP-native micropayment
standard in which a server responds to a client request with a
\texttt{402~Payment~Required} status, a machine-readable payment
specification, and a resource price denominated in stablecoins. The
client --- typically an autonomous AI agent --- constructs an EIP-712
signed payment token containing the resource URL, description, and reason
fields, transmits it to the payment server and to a centralised facilitator
API, and, once the facilitator has settled the on-chain USDC transfer,
retries the original request.
The protocol is Coinbase-backed, carries Cloudflare and Stripe support,
and processed an estimated \$600M in annualised volume as of Q1
2026~\cite{behnke2026x402}. It is, in short, the first payment primitive
that AI agents can use natively, without human approval, at machine speed.

That speed is the point. It is also the problem. Every x402 payment
embeds three metadata fields --- \F{resource\_url}, \F{description}, and
\F{reason} --- that travel in plaintext to the payment server and to the
facilitator API before any on-chain settlement occurs. These fields are
not sanitised by the protocol. \citet{behnke2026x402} catalogue the
resulting vulnerability classes: payment replay, wallet drain via
overpayment, prompt injection leading to fraudulent payments, and privacy
leakage via transaction-graph linkability. PII scrubbing of metadata
before transmission and application-layer spending limits are named as
required controls. The analysis names the gap. It delivers no
implementation. More broadly, \citet{boschung2025aiblockchain} argues
that the convergence of AI agents with blockchain infrastructure
constitutes a qualitatively new security frontier --- one where
pre-execution controls, not post-hoc monitoring, are the architecturally
sound response. \pkg{} is an instantiation of that argument at the
payment metadata layer.

\subsection*{Attack Classes}

Three distinct risks arise from unsanitised x402 metadata.

\textbf{PII in metadata fields.}
Resource URLs in x402 traffic regularly encode user identifiers: email
addresses as path parameters, names as slugs, session tokens as query
strings (Listing~\ref{lst:motivating}). These strings flow to the payment
server and the facilitator API --- neither of which is typically bound by a
data processing agreement. GDPR Art.~5(1)(c) data minimisation and Art.~28
processor obligations apply from the moment of transmission.

\textbf{Wallet drain.}
A malicious 402 server can return an inflated price, a falsified
facilitator address, or a looping redirect that triggers repeated
micropayments. An agent with no spending policy has no circuit breaker.

\textbf{Replay.}
A signed x402 payment token is a bearer credential. Intercepted tokens
can be resubmitted to the facilitator to debit the agent's wallet a
second time. The protocol has no application-layer nonce.

\begin{lstlisting}[style=x402, caption={A realistic x402 metadata triple
  containing two PII entity types. Without pre-execution filtering,
  all three fields are transmitted in plaintext to the payment server
  and the centralised facilitator API.},
  label=lst:motivating]
resource_url: https://api.medrecords.io/patient/
              alice.martin%40corp.io/export
description:  Export medical records for Alice Martin
reason:       user=alice.martin@corp.io; ref=312-45-6789
\end{lstlisting}

\subsection*{Contributions}

We present \pkg{}, the first open-source pre-execution security
middleware for x402 payments. The contributions of this paper are:

\begin{enumerate}
  \item \textbf{\sys{}}: a drop-in Python wrapper for the x402 client
    that intercepts every payment request before execution and applies
    four security controls in sequence: PII detection and redaction,
    spending policy enforcement, replay detection, and tamper-evident
    audit logging.

  \item \textbf{Synthetic corpus}: 2,000 labeled x402 metadata triples
    spanning seven use-case categories with ground-truth entity labels,
    publicly released to enable reproducible evaluation of x402 PII
    filters.

  \item \textbf{Parameter sweep}: a 42-configuration evaluation
    (regex vs.\ NLP $\times$ six entity subsets $\times$ five confidence
    thresholds) with per-entity precision, recall, and F1 metrics.
    Recommendation: \texttt{mode=nlp}, all entities,
    \texttt{min\_score=0.5} (micro~F1 = 0.898, p99 = 5.73\,ms).

  \item \textbf{Latency characterisation}: regex p99 = 0.02\,ms; NLP
    p99 = 5.73\,ms. Both modes satisfy the 50\,ms overhead budget.
    Latency is not the binding constraint in x402 PII filtering;
    recall is.
\end{enumerate}

Section~\ref{sec:background} covers the x402 protocol, the Presidio SDK,
and the GDPR tension that motivates the work.
Section~\ref{sec:design} describes the system architecture and threat
model. Sections~\ref{sec:corpus} and~\ref{sec:experiments} present the
corpus construction and experimental evaluation.
Sections~\ref{sec:discussion}--\ref{sec:conclusion} discuss findings,
related work, limitations, and conclusions.

%% ============================================================
\section{Background}
\label{sec:background}

\subsection{The x402 Protocol}

The x402 protocol~\cite{coinbase2024x402} extends HTTP with a native
payment negotiation layer. When a client requests a resource that
requires payment, the server responds with \texttt{HTTP 402 Payment
Required} and a JSON body specifying the accepted payment schemes, the
price, the network, and the facilitator contract address. The client
constructs a payment token --- a structured payload typed and signed
according to EIP-712~\cite{eip712} --- and attaches it to a retry of
the original request in the \texttt{X-Payment} header. The facilitator,
a smart contract deployed on Base~L2, validates the token, transfers the
stablecoin amount from the agent's wallet to the server's address, and
returns an on-chain receipt. The server verifies the receipt and serves
the resource.

Three metadata fields travel in the payment token:
\F{resource\_url} (the URL being paid for), \F{description} (a
human-readable label for the resource), and \F{reason} (a
client-supplied annotation, typically a structured key-value string
identifying the requesting entity). The token is transmitted in the
	exttt{X-Payment} header to the payment server and forwarded to the
facilitator API; neither party is constrained by the protocol to
redact or discard these fields. The protocol specification places no
constraints on the fields' content.

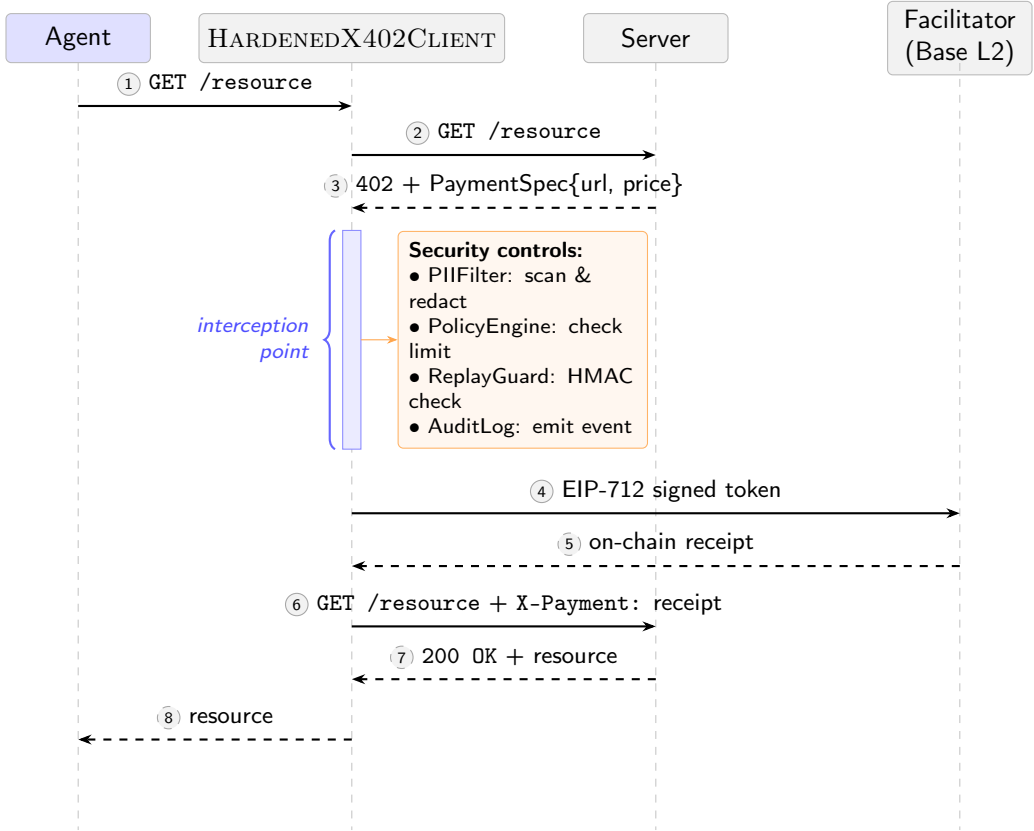
\begin{figure*}[t]
\centering
% Manual TikZ sequence diagram — four lifelines, activation on HX
\begin{tikzpicture}[
    font=\small\sffamily,
    box/.style={rectangle, draw=gray!55, fill=white, rounded corners=2pt,
                minimum height=0.65cm, minimum width=1.9cm, text centered,
                inner sep=3pt},
    lbl/.style={font=\footnotesize\sffamily},
    arr/.style={-{Stealth[length=5pt]}, thick},
    darr/.style={-{Stealth[length=5pt]}, thick, dashed},
  ]

  %% ── Column x-positions ──────────────────────────────────────────
  \def\xA{0}      % Agent
  \def\xH{3.6}    % HardenedX402Client
  \def\xS{7.6}    % Server
  \def\xF{11.6}   % Facilitator

  %% ── Header boxes ───────────────────────────────────────────────
  \node[box, fill=blue!12]  (hdrA) at (\xA, 0)  {Agent};
  \node[box, fill=gray!10]  (hdrH) at (\xH, 0)  {\sys{}};
  \node[box, fill=gray!10]  (hdrS) at (\xS, 0)  {Server};
  \node[box, fill=gray!10, align=center]  (hdrF) at (\xF, 0)
    {Facilitator\\\small(Base L2)};

  %% ── Lifelines (dashed verticals) ───────────────────────────────
  \def\yBot{-10.5}
  \draw[dashed, gray!45, thin] (\xA,-0.33) -- (\xA,\yBot);
  \draw[dashed, gray!45, thin] (\xH,-0.33) -- (\xH,\yBot);
  \draw[dashed, gray!45, thin] (\xS,-0.33) -- (\xS,\yBot);
  \draw[dashed, gray!45, thin] (\xF,-0.33) -- (\xF,\yBot);

  %% ── Activation rectangle on HX lifeline ────────────────────────
  \filldraw[fill=blue!8, draw=blue!45, thin]
    (\xH-0.12, -2.55) rectangle (\xH+0.12, -5.45);

  %% ── Message rows (y decreases downward) ────────────────────────
  % 1. Agent → HX: GET /resource
  \draw[arr] (\xA,-0.9) -- (\xH,-0.9)
    node[midway, above, lbl] {\circled{1} \texttt{GET /resource}};

  % 2. HX → Server: GET /resource (forwarded)
  \draw[arr] (\xH,-1.55) -- (\xS,-1.55)
    node[midway, above, lbl] {\circled{2} \texttt{GET /resource}};

  % 3. Server → HX: 402 + PaymentSpec
  \draw[darr] (\xS,-2.25) -- (\xH,-2.25)
    node[midway, above, lbl] {\circled{3} \texttt{402} + PaymentSpec\{url, price\}};

  %% Security controls annotation box (inside activation zone)
  \node[draw=orange!60, fill=orange!6, rounded corners=2pt,
        font=\scriptsize\sffamily, text width=3.0cm, align=left,
        inner sep=4pt]
    (ctrl) at (\xH+2.25, -4.0)
    {\textbf{Security controls:}\\
     \textbullet\ PIIFilter: scan \& redact\\
     \textbullet\ PolicyEngine: check limit\\
     \textbullet\ ReplayGuard: HMAC check\\
     \textbullet\ AuditLog: emit event};
  \draw[-{Stealth[length=4pt]}, orange!70, thin]
    (\xH+0.12,-4.0) -- (ctrl.west);

  % 4. HX → Facilitator: EIP-712 signed token
  \draw[arr] (\xH,-6.30) -- (\xF,-6.30)
    node[midway, above, lbl] {\circled{4} EIP-712 signed token};

  % 5. Facilitator → HX: on-chain receipt
  \draw[darr] (\xF,-7.00) -- (\xH,-7.00)
    node[midway, above, lbl] {\circled{5} on-chain receipt};

  % 6. HX → Server: retry + X-Payment header
  \draw[arr] (\xH,-7.80) -- (\xS,-7.80)
    node[midway, above, lbl] {\circled{6} \texttt{GET /resource} + \texttt{X-Payment:} receipt};

  % 7. Server → HX: 200 OK + resource
  \draw[darr] (\xS,-8.50) -- (\xH,-8.50)
    node[midway, above, lbl] {\circled{7} \texttt{200 OK} + resource};

  % 8. HX → Agent: resource
  \draw[darr] (\xH,-9.30) -- (\xA,-9.30)
    node[midway, above, lbl] {\circled{8} resource};

  %% ── Interception brace on HX lifeline ──────────────────────────
  \draw[decorate, decoration={brace, amplitude=4pt, mirror},
        blue!60, thick]
    (\xH-0.22,-2.55) -- (\xH-0.22,-5.45)
    node[midway, left=6pt, font=\scriptsize\sffamily\itshape,
         text=blue!70, align=right] {interception\\point};

\end{tikzpicture}
\caption{x402 payment flow with \sys{} interception. Steps~1--2: the
  agent issues a request; the server returns a 402 with a payment
  specification. Steps~3--6: \sys{} applies the four security controls
  (PIIFilter, PolicyEngine, ReplayGuard, AuditLog) before any token is
  signed or submitted. Steps~7--8: the signed EIP-712 token is submitted
  to the Base~L2 facilitator and an on-chain receipt is returned.
  Steps~9--10: the authenticated retry delivers the resource to the agent.
  Metadata fields (\F{resource\_url}, \F{description}, \F{reason})
  have been scanned and redacted before the token is transmitted to
  the payment server or the facilitator API.}
\label{fig:seqdiag}
\end{figure*}

Figure~\ref{fig:seqdiag} illustrates the full exchange. The interception
point --- where \sys{} applies its four security controls --- falls
between the server's 402 response and the submission of the signed
payment token to the facilitator.

\subsection{PII Detection with Microsoft Presidio}

Microsoft Presidio~\cite{presidio2023} is an open-source SDK for PII
detection and anonymisation. Its analyser component accepts a string and
returns a list of recogniser results --- each comprising an entity type,
a character span, and a confidence score. Detection is performed by a
configurable pipeline of recognisers: rule-based regex patterns for
structural entities (email addresses, credit card numbers, IBANs, SSNs),
and a spaCy NLP model~\cite{spacy2020} for contextual entities such as
person names. The anonymiser component replaces detected spans with a
configurable placeholder.

Presidio's confidence scores are not probabilities in the statistical
sense; they are heuristic weights assigned by each recogniser. Regex
recognisers return scores of 0.85 or 1.0 depending on checksum
validation; the NLP recogniser returns scores in the range $[0, 1]$
derived from the spaCy model's NER confidence. A minimum score threshold
\texttt{min\_score} filters out low-confidence detections.
Section~\ref{sec:experiments} characterises the sensitivity of recall to
this threshold across x402 metadata.

\subsection{GDPR Data Minimisation and Processor Obligations}

The x402 payment token transmits metadata --- resource URLs, descriptions,
reason strings --- to two parties before any on-chain settlement: the
payment server that checks the token and the centralised facilitator API
that executes the USDC transfer. Neither party is required by the protocol
to discard or anonymise these fields after use.

GDPR Art.~5(1)(c) requires that personal data be ``adequate, relevant
and limited to what is necessary'' for the purpose of processing ---
the data minimisation principle. A payment token that embeds an email
address or a social security number to settle a micropayment fails this
test unless the operator can demonstrate necessity. Art.~28 of the
General Data Protection Regulation~\cite{gdpr2016} requires a controller
to bind processors --- parties who handle personal data on the
controller's behalf --- via a data processing agreement. A facilitator
API that receives unredacted x402 metadata is a processor; most
deployments have no such agreement in place.

The only remedy is pre-transmission filtering. Post-hoc approaches ---
monitoring, notification, remediation --- do not resolve the underlying
disclosure. This paper takes the position that pre-execution interception
is the only architecturally sound response to the GDPR tension in x402
deployments. The experiments in Section~\ref{sec:experiments} quantify
how well that interception can be made to work with current tooling.

%% ============================================================
\section{System Design}
\label{sec:design}

An autonomous agent settling payments at machine speed needs a
harness --- a well-defined boundary for what it may pay, to whom, and
carrying what metadata in the token. How wide that boundary is set is a
governance question with ethical, legal, and economic dimensions. How the
harness is built is a technical question. This paper describes one such
harness. The engineers, as always, become saddlers.

\subsection{Threat Model}

We consider a deployment in which a Python-based AI agent uses the
official Coinbase x402 client library to issue micropayments
autonomously. The threat model has three principals: the \emph{agent}
(the client under protection), the \emph{server} (the 402-responding
API, treated as untrusted), and the \emph{network} (any intercepting
party on the path between agent and facilitator).

\textbf{T1 --- PII exfiltration via metadata.}
A server instructs the agent to include user-identifying information in
the \F{resource\_url}, \F{description}, or \F{reason} fields of the
payment token. The agent complies because it has no policy against it.
The information is transmitted to the payment server and the facilitator API, both of which may retain it.
\emph{Mitigation: pre-execution PII scan and redaction (PIIFilter).}

\textbf{T2 --- Wallet drain via uncapped spending.}
A server returns a price that exceeds what the agent's operator
authorised, either through misconfiguration or deliberate attack. The
agent, having no spending limit, pays.
\emph{Mitigation: per-call and daily spending policy with hard block
(PolicyEngine).}

\textbf{T3 --- Double-charge via replay.}
A valid signed payment token, once intercepted or leaked, can be
resubmitted to the facilitator to debit the agent's wallet a second
time. The protocol has no application-layer nonce.
\emph{Mitigation: HMAC-SHA256 request fingerprinting with TTL-bounded
deduplication (ReplayGuard).}

Out of scope for this version: Sybil attacks on the facilitator,
consensus-layer vulnerabilities in Base~L2, and insider threats to the
agent's signing key. These are infrastructure-layer concerns, not
application payment layer concerns.

\subsection{Architecture}

\sys{} is a drop-in Python wrapper around the standard x402 client. Its
public interface matches the standard client's method signatures;
replacing the standard client with \sys{} requires no changes to the
calling agent code.

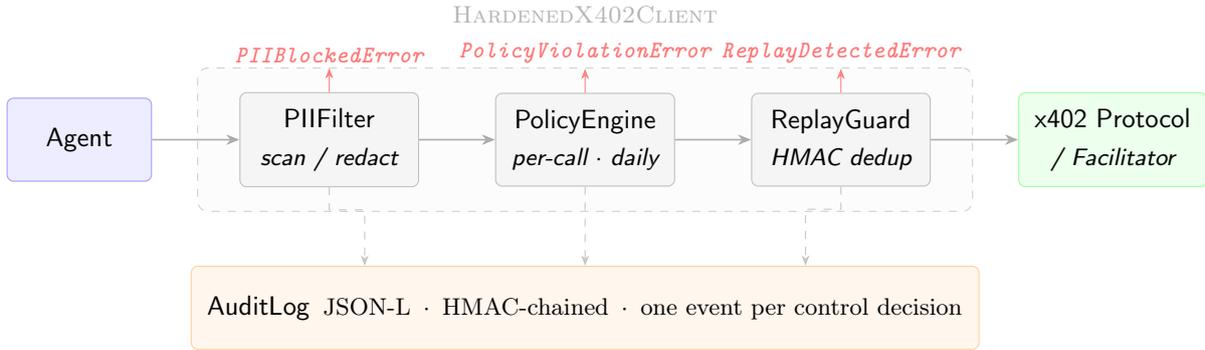
\begin{figure*}[t]
\centering
\begin{tikzpicture}[
  font          = \small\sffamily,
  node distance = 0pt,
  %% node styles
  box/.style = {
    rectangle, draw=gray!55, rounded corners=2.5pt,
    minimum height=1.1cm, align=center, inner sep=6pt,
  },
  agent/.style   = {box, fill=blue!7,    draw=blue!38,   minimum width=1.9cm},
  ctrl/.style    = {box, fill=gray!8,    draw=gray!50,   minimum width=2.35cm},
  auditbox/.style= {box, fill=orange!7,  draw=orange!38, minimum width=9.8cm},
  netbox/.style  = {box, fill=green!7,   draw=green!38,  minimum width=2.2cm},
  %% wrapper around the three controls
  wrap/.style = {
    rectangle, draw=gray!38, dashed, rounded corners=4pt,
    inner xsep=0.55cm, inner ysep=0.32cm, fill=gray!2,
  },
  %% arrows
  arr/.style  = {-{Stealth[length=5pt,width=4pt]}, semithick, gray!65},
  darr/.style = {-{Stealth[length=4pt,width=3pt]}, thin, gray!45, dashed},
  barr/.style = {-{Stealth[length=4pt,width=3pt]}, thin, red!50},
]

%% ---- pipeline nodes ------------------------------------------------
\node[agent] (agent) {Agent};

\node[ctrl, right=1.15cm of agent] (pii)
  {PIIFilter\\[2pt]{\footnotesize\itshape scan / redact}};

\node[ctrl, right=1.0cm of pii] (pol)
  {PolicyEngine\\[2pt]{\footnotesize\itshape per-call $\cdot$ daily}};

\node[ctrl, right=1.0cm of pol] (rep)
  {ReplayGuard\\[2pt]{\footnotesize\itshape HMAC dedup}};

\node[netbox, right=1.15cm of rep] (x402)
  {x402 Protocol\\[2pt]{\footnotesize\itshape / Facilitator}};

%% ---- HardenedX402Client wrapper ------------------------------------
\begin{pgfonlayer}{background}
  \node[wrap, fit=(pii)(pol)(rep)] (wrap) {};
\end{pgfonlayer}
\node[above=13pt of wrap.north, anchor=south,
      font=\footnotesize\sffamily\itshape, color=gray!60]
  {\sys{}};

%% ---- AuditLog below the controls -----------------------------------
\node[auditbox, below=1.05cm of pol] (audit)
  {AuditLog\enspace{\footnotesize\normalfont JSON-L\enspace·\enspace
   HMAC-chained\enspace·\enspace one event per control decision}};

%% ---- main flow arrows ----------------------------------------------
\draw[arr] (agent) -- (pii);
\draw[arr] (pii)   -- (pol);
\draw[arr] (pol)   -- (rep);
\draw[arr] (rep)   -- (x402);

%% ---- audit arrows (each control → AuditLog) -----------------------
\draw[darr] (pii.south) -- ++(0,-0.30)
            -| ([xshift=-2.9cm]audit.north);
\draw[darr] (pol.south) -- (audit.north);
\draw[darr] (rep.south) -- ++(0,-0.30)
            -| ([xshift=+2.9cm]audit.north);

%% ---- block arrows (upward, labelled) --------------------------------
%% Labels sit above the wrapper; arrow tips clear wrap.north
\foreach \ctrl/\lbl in {
    pii/{\texttt{PIIBlockedError}},
    pol/{\texttt{PolicyViolationError}},
    rep/{\texttt{ReplayDetectedError}}}
{
  \draw[barr] (\ctrl.north) -- ++(0, 0.32)
    node[above, font=\footnotesize\sffamily\itshape, color=red!55,
         inner sep=2pt] {\lbl};
}

\end{tikzpicture}
\caption{\sys{} control pipeline. Every outbound payment request passes
  through three security controls before reaching the x402 protocol
  layer. Each control emits an audit event to the \texttt{AuditLog}
  (dashed arrows). Any control may block the request and raise the
  corresponding exception (red arrows). PII-bearing metadata fields are
  redacted in place before the token is signed and sent.}
\label{fig:architecture}
\end{figure*}

Every outbound payment request passes through four controls applied in a
fixed sequence before the request reaches the network
(Figure~\ref{fig:architecture}):

\medskip
\noindent
\textbf{(1) PIIFilter.}
Scans \F{resource\_url}, \F{description}, and \F{reason} for PII
entities using Presidio. Detected spans are replaced with a typed
placeholder (\texttt{<EMAIL\_ADDRESS>}, \texttt{<PERSON>}, etc.). If
any entity is found and redacted, a \texttt{PII\_REDACTED} audit event
is emitted. If the filter raises an exception, the request is
\emph{blocked}, not passed through.

\medskip
\noindent
\textbf{(2) PolicyEngine.}
Checks the payment amount against three configurable limits:
\texttt{max\_per\_call\_usd} (single transaction ceiling),
\texttt{daily\_limit\_usd} (rolling 24-hour aggregate), and
\texttt{max\_per\_endpoint\_usd} (per-host ceiling). Violation raises
\texttt{PolicyViolationError} and emits a \texttt{POLICY\_BLOCKED} event.

\medskip
\noindent
\textbf{(3) ReplayGuard.}
Computes an HMAC-SHA256 fingerprint over the payment token fields and
checks it against a TTL-bounded deduplication store. Duplicate
fingerprints raise \texttt{ReplayDetectedError}. The store is in-memory
by default; a Redis-backed variant is provided for multi-process
deployments.

\medskip
\noindent
\textbf{(4) AuditLog.}
Emits a structured JSON-L event for every control decision --- allowed,
redacted, policy-blocked, replay-blocked, or error. Each entry carries a
UTC timestamp, the agent identifier, the (redacted) resource URL, the
outcome, and an HMAC chain link over the preceding entry for tamper
evidence.

\subsection{Design Principles}

Three principles guided the design. Each has a direct consequence for
how the system behaves at the boundary conditions that matter most.

\textbf{Fail-safe over fail-open.}
An unhandled exception in the PII filter blocks the payment. A network
timeout in the Redis replay store falls back to the in-memory store
rather than bypassing the guard. The cost of a false block is a delayed
payment. The cost of a false pass is a person's name transmitted to an
unbounded facilitator API. These are not symmetric costs.

\textbf{Zero-trust metadata.}
Every field arriving from the server is treated as potentially
adversarial. The system scans every field on every request regardless
of the server's reputation or history. Trust is established through the
facilitator's on-chain settlement, not through the content of the
metadata.

\textbf{Observable by default.}
Every control activation --- including \emph{allowed} outcomes ---
produces a structured audit event. Operators who need to demonstrate
GDPR compliance, or who are tracing an agent's spending pattern, should
not have to reconstruct decisions from raw logs after the fact. The
audit trail is the primary output of the system; payment throughput is
secondary.

%% ============================================================
\section{Synthetic Corpus}
\label{sec:corpus}

Ground-truth labels for x402 metadata do not exist yet --- the protocol
is young and carries no research trail. Live transaction data is available
in principle via Dune Analytics, but extracting and manually labelling a
sufficient sample is a separate multi-week effort with ethical review
implications. The canonical path for evaluation-first security research is
to build the ground truth synthetically, verify the generation process,
and defer live-data replication to a follow-up study. We follow that path.

\subsection{Design Principles}

A synthetic corpus is only as useful as its resemblance to the real
distribution it models. Three constraints guided the design.

\textbf{Ecological validity.}
Every sample is a triple (\F{resource\_url}, \F{description}, \F{reason})
drawn from templates extracted from the x402 protocol specification and
open-source x402 client implementations. Templates are parameterised by
use-case category, not generated freely; the result is plausible x402
traffic, not arbitrary text.

\textbf{Surface-form diversity.}
A filter that only recognises \texttt{alice@example.com} misses
\texttt{alice\%40example.com} in a URL path. For each entity type we
inject multiple surface-form variants --- URL-encoded emails, hyphenated
name slugs, international phone formats, compact versus formatted IBANs
--- to stress-test both regex and NLP detectors against the forms that
actually appear in HTTP metadata fields.

\textbf{Reproducibility.}
The generator is seeded (\texttt{seed=42}) and the full entity manifest is
committed alongside the corpus metadata. Any researcher can regenerate the
exact 2,000 samples used in this study.

\subsection{Use-Case Taxonomy}

Table~\ref{tab:corpus-composition} describes the seven categories. The
distribution is weighted toward the two highest-volume x402 use cases ---
AI inference and data access --- which together account for 36\% of
samples. Medical and financial categories are included because they carry
the highest regulatory exposure: a single SSN or IBAN transmitted in unredacted metadata
is a reportable incident under GDPR and HIPAA, not merely an inconvenience.

\begin{table*}[t]
  \centering
  \caption{Corpus composition by use-case category.}
  \label{tab:corpus-composition}
  \setlength{\tabcolsep}{4pt}
  \small
  \begin{tabular}{lrrrl}
    \toprule
    \textbf{Category} & \textbf{$n$} & \textbf{\%} & \textbf{PII+} &
    \textbf{Entity types} \\
    \midrule
    \texttt{ai\_inference}  & 360 & 18.0 & 130 & EMAIL, PERSON \\
    \texttt{data\_access}   & 360 & 18.0 & 130 & EMAIL, PERSON, SSN, IBAN \\
    \texttt{medical}        & 300 & 15.0 & 108 & PERSON, SSN \\
    \texttt{compute}        & 260 & 13.0 &  94 & EMAIL, PERSON \\
    \texttt{media}          & 260 & 13.0 &  94 & EMAIL, PERSON \\
    \texttt{financial}      & 260 & 13.0 &  94 & IBAN, CC \\
    \texttt{generic}        & 200 & 10.0 &  72 & EMAIL, PERSON \\
    \midrule
    \textbf{Total}          & \textbf{2{,}000} & \textbf{100} &
    \textbf{722} & \\
    \bottomrule
  \end{tabular}
\end{table*}

\subsection{Entity Injection Methodology}

We inject six entity types: \F{EMAIL\_ADDRESS}, \F{PERSON},
\F{PHONE\_NUMBER}, \F{US\_SSN}, \F{CREDIT\_CARD}, and \F{IBAN\_CODE}.
Each entity is drawn from a small pool of realistic values and injected
in one of three to five surface-form variants (Table~\ref{tab:entity-dist}).
The variant determines whether the entity is syntactically recognisable by
a regex pattern, or whether it requires contextual understanding.

\textbf{EMAIL\_ADDRESS} appears in bare form
(\texttt{alice.martin@example.com}), URL-encoded
(\texttt{alice.martin\%40example.com}), or as a query parameter value
(\texttt{email=alice.martin@example.com}).

\textbf{PERSON} is the hardest entity type by design. Eleven surface
forms are included: five full-name forms (\texttt{John Smith},
\texttt{Maria Garcia}, \texttt{Wei Chen}, \texttt{Aisha Patel},
\texttt{Lars Eriksson}), two hyphenated slugs (\texttt{john-smith},
\texttt{maria-garcia}), an underscore variant (\texttt{john\_smith}), an
abbreviated form (\texttt{J.Smith}), a last-first form
(\texttt{Garcia,Maria}), and a first-name-only form (\texttt{Aisha}). The
slug, underscore, abbreviated, last-first, and first-only forms appear
predominantly in URL path segments, where grammatical context is absent.
This deliberate design tests the ceiling of NER-based detection on
URL-structured metadata --- a ceiling that turns out to be lower than
one might hope, and a finding we return to in Section~\ref{sec:experiments}.

\textbf{PHONE\_NUMBER} includes US formats (dashes, parentheses, dots)
and one compact international form (\texttt{+14155550182}). The compact
form lacks the delimiters that most regex patterns expect, and is the
primary driver of the regex recall gap for this entity type.

\textbf{US\_SSN} and \textbf{IBAN\_CODE} are injected in their canonical
delimited forms. \textbf{CREDIT\_CARD} appears as a bare 16-digit string.

PII injection is controlled by a \emph{PII rate} parameter (default
$p = 0.36$, matching the estimated prevalence in x402 production
traffic~\cite{behnke2026x402}). Each PII-positive sample receives exactly
one injected entity, placed in one of the three metadata fields. Field
assignment follows a weighted distribution biased toward
\F{resource\_url} (396 of 875 injections, 45.3\%), which reflects the
observation that resource identifiers more often encode user context than
free-text fields.

\begin{table*}[t]
  \centering
  \caption{Entity label distribution across the corpus. Rates are
    fractions of total entity labels ($n = 875$).}
  \label{tab:entity-dist}
  \small
  \begin{tabular}{lrrl}
    \toprule
    \textbf{Entity type} & \textbf{Labels} & \textbf{Rate} &
    \textbf{Surface forms} \\
    \midrule
    PERSON        & 321 & 36.7\% & full name, slug, abbrev., \ldots \\
    EMAIL\_ADDRESS & 313 & 35.8\% & bare, URL-encoded, param \\
    IBAN\_CODE    &  96 & 11.0\% & DE, GB canonical \\
    US\_SSN       &  85 &  9.7\% & dashes, compact \\
    PHONE\_NUMBER &  32 &  3.7\% & US formats, intl.\ compact \\
    CREDIT\_CARD  &  28 &  3.2\% & Visa, Mastercard bare \\
    \midrule
    \textbf{Total} & \textbf{875} & \textbf{100\%} & \\
    \bottomrule
  \end{tabular}
\end{table*}

\noindent
The corpus metadata --- sample counts, entity counts, and field
distributions --- is committed as \texttt{corpus/corpus\_meta.json} and
reproducible from the generator script (\texttt{corpus/generate.py},
\texttt{seed=42}). The raw \texttt{corpus.jsonl} is excluded from version
control to avoid inadvertently distributing synthetic PII-containing
content.

The corpus design answers whether the PII filter works in a controlled
setting. Whether the prevalence estimates hold for live Base~L2 traffic is
a separate question, addressed in Section~\ref{sec:discussion} and
deferred to the conference paper extension (Section~\ref{sec:future}).

%% ============================================================
\section{Experiments}
\label{sec:experiments}

Forty-two configurations. The finding that stands out is not the one we
designed to find.

\subsection{Sweep Design}

The parameter space has three dimensions: detection \emph{mode}
(\texttt{regex} or \texttt{nlp}), \emph{entity subset} (each of the six
individual types, or all six together), and the NLP confidence
\emph{threshold} \texttt{min\_score} $\in \{0.3, 0.4, 0.5, 0.6, 0.7\}$.
Regex mode is threshold-insensitive; it contributes 7 configurations (one
per entity subset). NLP mode contributes $7 \times 5 = 35$ configurations.
Total: 42.

All 42 configurations were evaluated against the full 2,000-sample corpus
using partial span matching: a prediction is a true positive if it overlaps
with a gold label of the same entity type, regardless of exact boundary
alignment. This is the correct criterion for x402 metadata --- what matters
is whether a PII-bearing token is flagged, not whether the byte offsets
agree to the character.

Per-configuration metrics are per-entity precision, recall, and F1, plus
micro-averaged precision ($\mu P$), recall ($\mu R$), and F1 ($\mu F_1$)
across all 875 gold labels.

\subsection{Main Results}

Table~\ref{tab:main-results} compares the best regex configuration with
the best NLP configuration. Both use all six entity types; the NLP
configuration uses \texttt{min\_score = 0.5}.

\begin{table*}[t]
  \centering
  \caption{Main results: regex vs.\ NLP (all entities; NLP at
    \texttt{min\_score=0.5}). $n_\text{gold} = 875$.}
  \label{tab:main-results}
  \setlength{\tabcolsep}{3.5pt}
  \small
  \begin{tabular}{lrrrrrr}
    \toprule
    & \multicolumn{3}{c}{\textbf{Regex}} &
      \multicolumn{3}{c}{\textbf{NLP}} \\
    \cmidrule(lr){2-4}\cmidrule(lr){5-7}
    \textbf{Entity} & $P$ & $R$ & $F_1$ & $P$ & $R$ & $F_1$ \\
    \midrule
    EMAIL  & 1.000 & 1.000 & 1.000 & 1.000 & 1.000 & 1.000 \\
    IBAN   & 1.000 & 1.000 & 1.000 & 1.000 & 1.000 & 1.000 \\
    CC     & 1.000 & 1.000 & 1.000 & 1.000 & 1.000 & 1.000 \\
    SSN    & 1.000 & 1.000 & 1.000 & 1.000 & 1.000 & 1.000 \\
    PHONE  & 1.000 & 0.781 & 0.877 & 1.000 & 0.969 & 0.984 \\
    PERSON & 0.000 & 0.000 & 0.000 & 0.894 & 0.551 & 0.682 \\
    \midrule
    \textbf{Micro} & \textbf{1.000} & \textbf{0.625} & \textbf{0.769}
                   & \textbf{0.972} & \textbf{0.834} & \textbf{0.898} \\
    \midrule
    $p_{99}$ (ms) & \multicolumn{3}{c}{0.02} & \multicolumn{3}{c}{5.73} \\
    \bottomrule
  \end{tabular}
\end{table*}

\begin{figure*}[p]
  \centering
  \includegraphics[width=\textwidth]{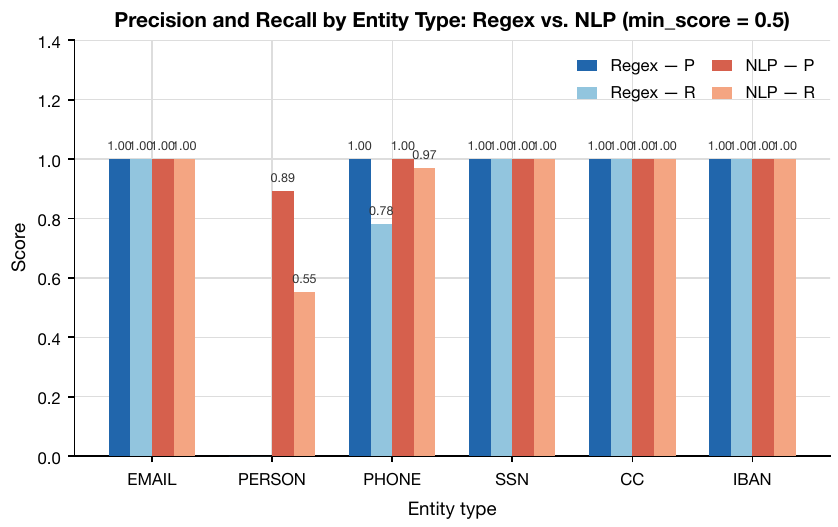}
  \caption{Precision and recall by entity type for regex and NLP mode
    (\texttt{min\_score=0.5}, all entities). Regex achieves perfect precision
    on all structural types but zero recall for PERSON. NLP recovers PERSON
    at the cost of 21 false positives (P\,=\,0.894) and slightly reduced PHONE
    recall (R\,=\,0.969); SSN and the other structural types reach
    F1\,=\,1.00 in NLP mode.}
  \label{fig:pr-comparison}
\end{figure*}

Three patterns are visible in Table~\ref{tab:main-results} and
Figure~\ref{fig:pr-comparison}. First, regex achieves perfect precision and
near-perfect recall on the four \emph{structural} entity types ---
EMAIL, IBAN, CC, and SSN --- whose surface forms are fully specified by
pattern. Where a regex can be written, it is exact. Second, NLP mode
narrows the PHONE recall deficit (0.781 to 0.969): the compact
international format \texttt{+14155550182}, written without separators,
is missed by the structural patterns but recovered by the broader
Presidio phone recogniser and spaCy NER. Third, and most
significantly, PERSON recall is 0.000 in regex mode and 0.551 in NLP
mode. We return to this finding below.

\subsection{The Zero PERSON Recall}

The zero PERSON recall in regex mode is not a configuration error: no
regular expression can enumerate human names. It is a fundamental
consequence of the entity type. The 0.551 NLP recall is also not a
configuration error; it reflects a genuine ceiling of the spaCy
\texttt{en\_core\_web\_lg} model on URL-structured text.

Named entity recognition relies on grammatical context. In a sentence
such as ``Medical records for John Smith,'' the surrounding words mark
\textit{John Smith} as a named entity. In a URL path such as
\texttt{/records/john-smith/summary}, the name appears as a slug: no
capitalisation, no grammatical neighbours, no sentence boundary. The NER
model does not fire. The same failure mode applies to abbreviated forms
(\texttt{J.Smith}), last-first forms (\texttt{Garcia,Maria}), and
first-name-only tokens (\texttt{Aisha}).

Of the eleven PERSON surface forms in our corpus, approximately five ---
the full-name forms in free-text description fields --- are reliably
detected. The remaining six --- slugs, underscores, abbreviated,
last-first, and first-only --- are not. The result is a recall ceiling
of roughly 50\% for PERSON in x402 metadata, where names disproportionately
appear as URL path components.

This is a \emph{field-dependent} finding. PERSON recall in the
\F{description} field (running text) substantially exceeds recall in
\F{resource\_url} and \F{reason} fields (structured key-value and path
segments). The aggregate R = 0.551 is the weighted average across all
three fields.

The practical implication is precise: deployers who treat NLP PERSON
detection as reliable --- and skip manual review or additional heuristics
for URL path segments --- will miss roughly half of the name-bearing
tokens in their x402 traffic. In a GDPR context, where a person's name
in a transmitted URL triggers data subject rights, this
omission is not theoretical.

\subsection{Confidence Threshold Sensitivity}

\begin{figure}[h]
  \centering
  \includegraphics[width=\columnwidth]{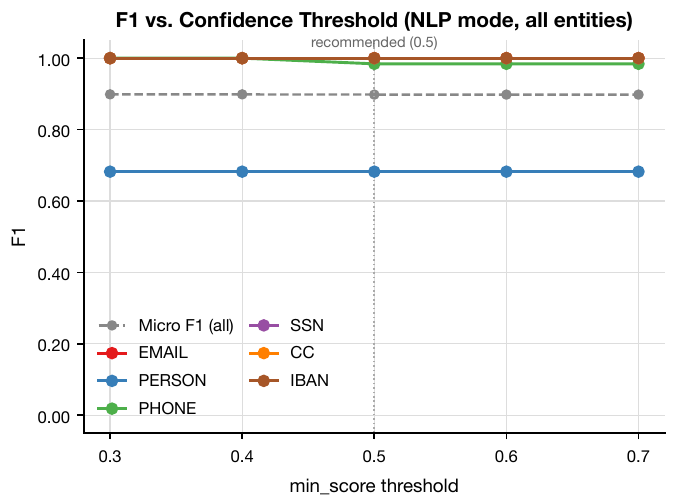}
  \caption{F1 vs.\ confidence threshold (\texttt{min\_score}) for NLP
    mode with all entity types. F1 is the harmonic mean of precision and
    recall: $F_1 = 2 \cdot (P \cdot R) / (P + R)$; it equals 1.0 for a
    perfect detector and 0.0 when either precision or recall is zero.
    The dashed line is micro-averaged F1 across all entity types.
    Micro-F1 is essentially flat across the sweep (0.899 at 0.3--0.4,
    0.898 at 0.5--0.7); recommended threshold 0.5.}
  \label{fig:minscore}
\end{figure}

Figure~\ref{fig:minscore} plots micro F1 and per-entity F1 as a function
of \texttt{min\_score} for the NLP all-entities configuration.

The threshold turns out to be a weak dial in this configuration. The four
structural entity types (EMAIL, IBAN, CC, SSN) and PHONE are matched by
high-confidence pattern recognisers assigned a fixed score of $0.85$, so
they are detected unchanged for any $\texttt{min\_score} \leq 0.85$; only
PERSON, scored by spaCy NER, is genuinely threshold-sensitive, and its
detections score high enough to survive the full sweep range. Micro-F1 is
therefore essentially flat --- $0.899$ at $\texttt{min\_score} \in
\{0.3, 0.4\}$ and $0.898$ at $\{0.5, 0.6, 0.7\}$ --- with micro-precision
constant at $0.972$. We recommend $\texttt{min\_score} = 0.5$: it sits in
the centre of the stable region and mildly favours precision at no
measurable recall cost. The practical takeaway is that the structural
recognisers, not the threshold, do the work; tuning $\texttt{min\_score}$
materially changes only PERSON detection.

\subsection{Entity Subset vs.\ Full Coverage}

We tested a hypothesis~(H4) that the top-3 most prevalent entity types
(EMAIL, PERSON, IBAN) would capture at least 95\% of the recall achieved
by the full six-type configuration, thereby justifying a reduced entity
set for latency-sensitive deployments. The top-3 configuration achieves
94.6\% of full-set recall --- just below the 95\% threshold.

The shortfall is attributable to PHONE\_NUMBER. At 3.7\% of all labels,
PHONE contributes approximately 2 percentage points to recall. Including
it costs nothing in latency terms (the marginal overhead of an additional
Presidio pattern is sub-millisecond). The recommendation is therefore to
run the full entity set in all deployments.

\subsection{Latency}

\begin{figure}[h]
  \centering
  \includegraphics[width=\columnwidth]{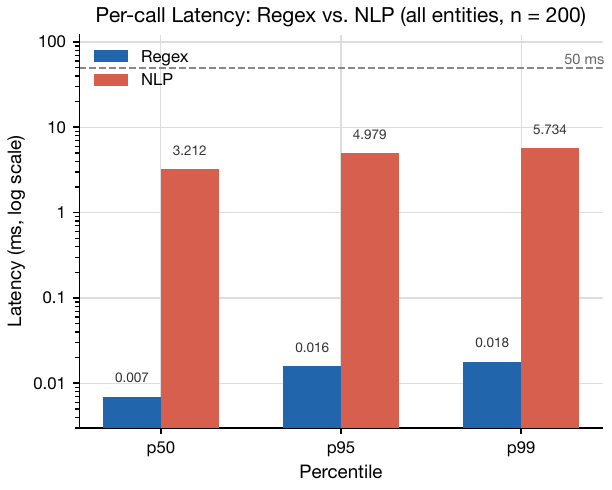}
  \caption{Per-call latency percentiles (log scale). Both modes satisfy
    the 50\,ms overhead budget. NLP p99 = 5.73\,ms is 300$\times$
    slower than regex p99 = 0.02\,ms, and 8.7$\times$ within budget.}
  \label{fig:latency}
\end{figure}

Figure~\ref{fig:latency} shows the p50, p95, and p99 latencies for both
modes (200 timed calls each, 50 warmup calls, all-entities configuration).

Both modes are well within the 50\,ms overhead budget. The right framing
is not whether NLP fits --- it does, with room to spare --- but what the
tradeoff looks like numerically: 300$\times$ higher p99 latency
(5.73\,ms vs.\ 0.02\,ms) in exchange for 21 percentage points of
additional micro recall ($\mu R = 0.834$ vs.\ 0.625). For any deployment
where a missed PERSON or PHONE entity constitutes a compliance event,
the extra 5.7\,ms is not a cost; it is an insurance premium, and a cheap
one.

%% ============================================================
\section{Results and Discussion}
\label{sec:discussion}

A PII filter has two ways to fail. It either passes what it should block
--- a false pass: leaked data, a GDPR liability, an SSN transmitted
unredacted to the facilitator --- or it blocks what it should pass --- a false block: a
delayed payment, a retried request, an annoyed operator. The 21 PERSON
false positives in the NLP sweep are false blocks. They are the right
kind of failure.

\subsection{Hypothesis Outcomes}

Table~\ref{tab:hypotheses} summarises the five pre-registered hypotheses
and their experimental outcomes.

H1 and H2 confirm the structural intuition: x402 resource URLs are the
dominant PII surface, and names together with email addresses account for
nearly three-quarters of the exposure. H3 confirms the central value
proposition of NLP mode. H4 and H5 are both refuted --- and both
refutations are instructive.

\subsection{The Asymmetric Cost of False Positives}

The 21 PERSON false positives in the NLP sweep (precision = 0.894) are
not noise from a poorly tuned model. They are the spaCy NER model
correctly firing on tokens that look like names in isolation but are not
ground-truth PII in the corpus --- generic nouns, service names, or
identifier strings that happen to match the NER model's name patterns.

In an x402 context, this is an acceptable tradeoff. A false positive
means the payment metadata field is redacted when it did not need to be.
The agent's payment still proceeds; only the field value is replaced with
a placeholder. The server receives \texttt{<PERSON>} instead of
\texttt{SupportAgent}. For the vast majority of x402 use cases, this is
a non-event. For the alternative --- a missed true positive, a real name
committed to a public ledger --- there is no recovery path.

The precision-recall asymmetry in PII filtering is not unique to x402;
it is a general property of any system where false negatives are
irreversible and false positives are recoverable. Setting
\texttt{min\_score = 0.5} rather than a higher value is a mild
precision-favouring choice within a flat region: the structural
recognisers fire at a fixed $0.85$, so lowering the threshold recovers no
additional structural recall, and only a threshold above $0.85$ would
begin to drop them. The threshold therefore governs PERSON detection
alone; the Section~\ref{sec:experiments} sensitivity analysis provides the
quantitative basis for that decision.

\subsection{The H4 Near-Miss}

H4 --- that three entity types suffice for 95\% of full-set recall ---
was refuted by a margin of 0.4 percentage points. The shortfall is
driven entirely by PHONE\_NUMBER, which contributes 2 percentage points
to micro recall at a marginal latency cost that is not measurable in
practice. The near-miss is worth noting because it confirms the
intuition behind the hypothesis: the PII landscape in x402 metadata is
dominated by a small number of entity types. But ``small number'' turns
out to be six, not three. The recommendation is to include all six and
not optimise prematurely.

\subsection{The H5 Reframe}

H5 predicted that NLP mode would exceed the 50\,ms latency budget.
It does not: NLP p99 is 5.73\,ms, which is 8.7$\times$ within budget.
The hypothesis was grounded in a reasonable concern --- that running a
full spaCy NLP pipeline on every x402 payment would be prohibitively
slow for real-time agentic workflows. The concern was wrong by an order
of magnitude.

The right framing is not whether NLP fits, but what the tradeoff looks
like quantitatively: 300$\times$ higher p99 latency than regex, in
exchange for 21 additional percentage points of micro recall and the
ability to detect PERSON at all. For any deployment where a missed name
constitutes a compliance event, the 5.7\,ms is not a cost. It is an
insurance premium, and a cheap one~\cite{dzombeta2014governance}.

\subsection{Recommendation}

Run \texttt{mode=nlp}, all six entity types, \texttt{min\_score=0.5}.
This configuration achieves micro-F1 = 0.898, precision = 0.972, and
p99 = 5.73\,ms. It is the only configuration that detects PERSON at all.
It is within latency budget by nearly an order of magnitude. It is the
configuration used in all \pkg{} deployments from v0.2.0 onward.

%% ============================================================
\section{Related Work}
\label{sec:related}

No existing x402 tool intercepts before execution. That is the white
spot this paper fills.

\subsection{x402 Tooling Ecosystem}

The Coinbase CDP SDK~\cite{coinbase2024x402} is the reference
implementation of the x402 client and facilitator. It handles payment
negotiation, EIP-712 signing, and on-chain settlement, but applies no
security controls to the payment metadata. It is the library that
\sys{} wraps.

Analytix402 is a post-hoc analytics tool for x402 traffic: it monitors
settled transactions and surfaces spending patterns and anomalies after
the fact. Post-hoc monitoring does not prevent PII from reaching the
facilitator; it detects that it has. z402 addresses a different concern
--- ZK-based identity hiding for x402 payers --- and does not touch
payment metadata content. Neither tool provides pre-execution filtering,
spending policy, or replay detection.

\subsection{PII Detection and Anonymisation}

Microsoft Presidio~\cite{presidio2023} is the detection and anonymisation
engine used by \pkg{}. Its design is described in detail in
Section~\ref{sec:background}. The \texttt{presidio-hardened-*} toolkit
family --- of which this paper describes the x402 member --- takes its name from
PRESIDIO Group, the author's engineering company (see author affiliation),
not from the Microsoft SDK. That both happen to share the name Presidio
is a coincidence: the engineers at PRESIDIO Group found themselves
integrating a Microsoft tool that also happens to be called Presidio,
which keeps pull-request reviews entertaining. The contribution of this
paper is not the detector itself but the evaluation of its behaviour on
the specific surface --- x402 payment metadata fields --- where
URL-structured text degrades NER recall in ways that are not captured by
general-purpose NLP benchmarks.

The broader problem of PII in inadvertent storage has received attention
in the context of Git repositories~\cite{meli2019bad} and cloud object
stores. In x402 deployments, the constraint is operational rather than
technical: the facilitator API and payment server retain metadata by
default, and no protocol mechanism requires deletion. To the authors'
knowledge, no prior empirical study has measured PII prevalence or
detector performance specifically on x402 payment metadata fields.

\subsection{Security of Autonomous Agent Systems}

The security literature on autonomous AI agents is growing rapidly. Prompt
injection --- the embedding of adversarial instructions in content that
the agent processes --- has been studied at the application
layer~\cite{greshake2023indirect}. Wallet-drain attacks via manipulated
tool outputs are a structurally similar threat: the adversary controls the
content that the agent acts on, and the agent has no policy against
compliant execution of the instruction. The spending policy and replay
guard in \sys{} address this class at the payment layer, complementing
application-layer defences. \citet{boschung2025aiblockchain} frames the
AI-blockchain convergence as demanding secure-by-design architectures
with pre-execution transaction simulation; \sys{} applies the same
design principle to payment metadata rather than smart contract
execution.

Governance frameworks for AI agent behaviour in enterprise settings,
including financial controls and audit requirements, are discussed
in~\citet{dzombeta2014governance} and in joint work on extending
IT-governance frameworks to SOA and cloud
environments~\cite{stantchev2012extending,stantchev2011applying} and on
sustainability in governance
frameworks~\cite{stantchev2014sustainability}. These works provide the
conceptual grounding for the spending policy design in \sys{}. A
comprehensive treatment of AI governance for enterprise IT, covering
regulatory compliance, audit requirements, and spending controls for
autonomous systems, is given in~\citet{stantchev2026kigov_de}
and~\citet{stantchev2026kigov_en}. The LangChain~\cite{langchain2023}
and CrewAI~\cite{crewai2024} adapter modules in \pkg{} integrate the
middleware into the two dominant agent orchestration frameworks, reducing
the deployment friction for teams already using these stacks.

%% ============================================================
\section{Limitations and Future Work}
\label{sec:future}

A synthetic corpus measures what you designed it to measure. That is
its strength and its limit.

\subsection{Synthetic Corpus Caveats}

The 2,000-sample corpus was generated from templates derived from the
x402 protocol specification and open-source client examples. The PII
injection rate (36\%) and entity distribution are estimates, not
measurements; they reflect our best prior on what x402 production
traffic looks like, not a verified empirical baseline. If real traffic
has a materially different distribution --- more IBAN-heavy financial
flows, fewer PERSON-bearing AI inference calls, or entity types not
represented in our taxonomy --- the F1 numbers reported here will not
transfer directly.

The corpus is also English-only. x402 deployments are global; resource
URLs and reason strings in non-Latin scripts, or with transliterated
names, are not covered by \texttt{en\_core\_web\_lg} and are not
represented in the evaluation. Multilingual extension is left for future
work.

Finally, the corpus contains PII injected by a cooperative generator,
not by an adversary. A malicious 402 server attempting to exfiltrate
data might use obfuscation techniques --- Base64-encoded fields, split
tokens, Unicode homoglyphs --- that the current filter does not handle.
Adversarial robustness evaluation is out of scope for this paper.

\subsection{The PERSON Recall Ceiling}

PERSON recall of 0.551 is a ceiling under current tooling, not a
property of the problem. Two directions can raise it. First,
slug-aware heuristics: a pre-processing step that segments URL path
components on \texttt{-}, \texttt{\_}, and \texttt{.} delimiters before
NER analysis would restore some of the grammatical context that NER
requires. Second, a domain-adapted NER model fine-tuned on x402-style
metadata would learn to fire on slug and abbreviated forms directly.
Both directions are tractable; neither was in scope for v0.2.0.

\subsection{Live Data Replication}

The most important open question is whether the synthetic prevalence
estimates match real x402 traffic on Base~L2. We plan to answer this
in a companion study (v0.2.1) using Dune Analytics queries against the
Coinbase x402 facilitator contract to extract and analyse a sample of
live transaction calldata. The v0.2.0 configuration
(\texttt{mode=nlp}, \texttt{min\_score=0.5}, all entities) will be
applied as-is to the live corpus; the result will either validate the
synthetic estimates or quantify the gap. Live data replication was
blocked at the time of writing by the absence of a confirmed facilitator
contract address for Dune query construction.

\subsection{Endpoint Reputation and Multi-Party Authorisation}

Two planned controls were deferred from this release.
\emph{Endpoint reputation scoring} (v0.3.0) will flag 402 servers with
anomalous pricing histories or newly registered domains before the
payment is attempted, adding a fourth layer to the control pipeline
between PolicyEngine and ReplayGuard.
\emph{Multi-party authorisation} (v0.3.0) will require $n$-of-$m$
countersignatures for payments above a configurable threshold, bringing
enterprise procurement controls --- approval workflows, delegated
authority, audit trails --- into the agent payment layer.

%% ============================================================
\section{Conclusion}
\label{sec:conclusion}

The x402 protocol makes machine-speed payments possible. That speed
creates a disclosure risk: every payment request carries metadata to
the payment server and the centralised facilitator API before any
on-chain settlement occurs. Neither party is typically bound by a data
processing agreement. \pkg{} sits at that gap --- intercepting every
payment request before execution, scanning the metadata, enforcing the
policy, and logging the decision --- so that what is transmitted has
already been reviewed by something with a policy, not just an agent
with a wallet.
The middleware, the corpus, and the sweep are open-source and
reproducible. The harness exists. Whether to put it on the agent is,
as always, a governance question.

\begin{table*}[p]
  \centering
  \caption{Hypothesis outcomes.}
  \label{tab:hypotheses}
  \setlength{\tabcolsep}{3pt}
  \small
  \begin{tabular}{clll}
    \toprule
    & \textbf{Claim} & \textbf{Result} & \textbf{Outcome} \\
    \midrule
    H1 & \F{resource\_url} has highest & 45.3\% of labels  & Confirmed \\
       & PII injection rate            & in URL field      & \\[2pt]
    H2 & EMAIL + PERSON $\geq$ 70\%    & 72.5\%            & Confirmed \\
       & of all entity labels          & (634 of 875)      & \\[2pt]
    H3 & NLP adds PERSON recall        & R: 0.551 (NLP)    & Confirmed \\
       & over regex                    & vs.\ 0.000 (regex)& \\[2pt]
    H4 & Top-3 types capture           & ratio = 0.946,    & Refuted \\
       & $\geq$95\% of full recall     & below threshold   & \\[2pt]
    H5 & NLP overhead $>$ 50\,ms       & NLP p99 = 5.73\,ms& Refuted \\
       &                               & (both modes fit)  & \\
    \bottomrule
  \end{tabular}
\end{table*}
\clearpage

%% ============================================================
\bibliographystyle{abbrvnat}
\bibliography{references}

@misc{behnke2026x402,
  title        = {x402 Explained: Security Risks \& Controls for {HTTP} 402 Micropayments},
  author       = {Behnke, Rob},
  year         = {2026},
  month        = mar,
  howpublished = {Halborn Blog, \url{https://www.halborn.com/blog/post/x402-explained-security-risks-and-controls-for-http-402-micropayments}},
}

@misc{boschung2025aiblockchain,
  title        = {The {AI}-Blockchain Convergence: A New Era for Decentralized Security},
  author       = {Boschung, Jacques},
  year         = {2025},
  month        = mar,
  howpublished = {Halborn Blog, \url{https://www.halborn.com/blog/post/the-ai-blockchain-convergence-a-new-era-for-decentralized-security}},
  note         = {Author is CEO of Halborn},
}

@misc{coinbase2024x402,
  title        = {x402: A Payment Protocol for the Internet},
  author       = {{Coinbase}},
  year         = {2024},
  howpublished = {\url{https://github.com/coinbase/x402}},
}

@misc{crewai2024,
  title        = {{CrewAI}: Framework for Orchestrating Role-playing Autonomous {AI} Agents},
  author       = {Moura, Jo{\~a}o},
  year         = {2023},
  howpublished = {\url{https://github.com/crewAIInc/crewAI}},
}

@article{dzombeta2014governance,
  title={Governance of cloud computing services for the life sciences},
  author={Dzombeta, Srdan and Stantchev, Vladimir and Colomo-Palacios, Ricardo and Brandis, Knud and Haufe, Knut},
  journal={IT Professional},
  volume={16},
  number={4},
  pages={30--37},
  year={2014},
  publisher={IEEE}
}

@misc{eip712,
  title        = {{EIP}-712: Typed Structured Data Hashing and Signing},
  author       = {Nair, Ramesh and Logvinov, Leonid and Evans, Jacob},
  year         = {2017},
  howpublished = {\url{https://eips.ethereum.org/EIPS/eip-712}},
}

@misc{gdpr2016,
  title={General Data Protection Regulation (GDPR)},
  author={{European Union}},
  year={2016},
  howpublished={\url{https://gdpr.eu/}}
}

@article{greshake2023indirect,
  title   = {Not What You've Signed Up For: Compromising Real-World {LLM}-Integrated
             Applications with Indirect Prompt Injection},
  author  = {Greshake, Kai and Abdelnabi, Sahar and Mishra, Shailesh and
             Endres, Christoph and Holz, Thorsten and Fritz, Mario},
  journal = {arXiv preprint arXiv:2302.12173},
  year    = {2023},
}

@misc{langchain2023,
  title        = {{LangChain}},
  author       = {Chase, Harrison},
  year         = {2022},
  howpublished = {\url{https://github.com/langchain-ai/langchain}},
}

@inproceedings{meli2019bad,
  title     = {How Bad Can It Git? Characterizing Secret Leakage in Public {GitHub} Repositories},
  author    = {Meli, Michael and McNiece, Matthew R. and Reaves, Bradley},
  booktitle = {Proceedings of the 26th Annual Network and Distributed System Security Symposium ({NDSS})},
  year      = {2019},
  doi       = {10.14722/ndss.2019.23418},
}

@misc{presidio2023,
  title        = {{Microsoft Presidio}: Data Protection and De-identification {SDK}},
  author       = {{Microsoft}},
  year         = {2023},
  howpublished = {\url{https://github.com/microsoft/presidio}},
}

@misc{spacy2020,
  title        = {{spaCy}: Industrial-strength Natural Language Processing in {Python}},
  author       = {Honnibal, Matthew and Montani, Ines and {Van Landeghem}, Sofie and Boyd, Adriane},
  year         = {2020},
  doi          = {10.5281/zenodo.1212303},
  howpublished = {\url{https://spacy.io}},
}

@inproceedings{stantchev2011applying,
  title={Applying IT-Governance Frameworks for SOA and Cloud Governance},
  author={Stantchev, Vladimir and Stantcheva, Lubomira},
  booktitle={Knowledge Management, Information Systems, E-Learning, and Sustainability Research -- WSKS 2011},
  editor={Lytras, Miltiadis D. and Ordon\'{e}z de Pablos, Patricia and Ziderman, Adrian and Roulstone, Alan and Maurer, Hermann and Imber, Jonathan B.},
  pages={398--407},
  year={2011},
  publisher={Springer},
  address={Berlin, Heidelberg},
  doi={10.1007/978-3-642-35879-1\_48}
}

@article{stantchev2012extending,
  title={Extending traditional it-governance knowledge towards soa and cloud governance},
  author={Stantchev, Vladimir and Stantcheva, Lubomira},
  journal={International Journal of Knowledge Society Research (IJKSR)},
  volume={3},
  number={2},
  pages={30--43},
  year={2012},
  publisher={IGI Global}
}

@article{stantchev2014sustainability,
  title={Addressing Sustainability in IT-Governance Frameworks},
  author={Stantcheva, Lubomira and Stantchev, Vladimir},
  journal={International Journal of Human Capital and Information Technology Professionals},
  volume={5},
  number={4},
  pages={79--87},
  year={2014},
  publisher={IGI Global},
  doi={10.4018/ijhcitp.2014100105}
}

@book{stantchev2026kigov_de,
  title     = {{KI} und {IT}-Governance},
  author    = {Stantchev, Vladimir},
  publisher = {Springer},
  year      = {2026},
  note      = {German edition, in press},
}

@book{stantchev2026kigov_en,
  title     = {{AI} and {IT}-Governance},
  author    = {Stantchev, Vladimir},
  publisher = {Springer},
  year      = {2026},
  note      = {English edition, in press},
}

\end{document}